\documentclass[a4paper,10pt]{aa}
\usepackage{natbib}
\usepackage{epsfig}
\begin{document}
\newcommand{\Msun}{R$_{\sun}$\ }
\newcommand{\Lsun}{L$_{\sun}$\ }
\newcommand{\Mstar}{R$_{\star}$\ }
\newcommand{\Lstar}{L$_{\star}$\ }
\authorrunning{E. Whelan}
\title{Paschen beta emission as a tracer of outflow activity from T-Tauri stars, as compared to Optical Forbidden Emission}
\titlerunning{Paschen beta emission as a tracer of outflow activity}
\date{Received date ;accepted date}
\author{E.T. Whelan \inst{1}
\ T.P. Ray \inst{1} \ C.J. Davis \inst{2}}
\institute{ Dublin Institute for Advanced Studies, 5 Merrion Square, Dublin 2, Ireland 
\and  Joint Astronomy Centre, 660 North A'ohoku Place, University Park, Hilo, Hawaii 96720, USA}
\date{Received date; accepted date}
\abstract{The Paschen beta (1.2822 $\mu$m) emission line found in the near-infrared spectra of T-Tauri stars (TTSs) is believed to trace the accretion of material onto the central star. We present spectroscopic data which suggests that this may not always be the case. The technique of spectro-astrometry is used by us to measure positional displacements in the Pa$\beta$ emission from four T-Tauri stars, namely DG Tau, V536 Aql, LkH$\alpha$ 321 and RW Aur. We also observed the optical forbidden emission from these sources, for example the [SII]$\lambda\lambda$6716, 6731, [OI]$\lambda\lambda$6300, 6363 and [NII]$\lambda\lambda$6548, 6583 lines. Forbidden emission lines are formed in the outflows that accompany the evolution of protostars and so are ideal to use as a comparison to confirm that the measured offsets in the Pa$\beta$ emission are indeed due to outflowing material. Models based on the magnetospheric accretion theory have been the most successful to date in explaining the origin of atomic hydrogen emission lines. Yet we see that the line profiles of the sources showing displacement in their Pa$\beta$ emission all have features that the magnetospheric accretion model has so far failed to explain, such as broad full width half maxima, large wings and an absence of red shifted absorption features. The failure of the models to explain the presence of large extended wings in the line profiles is particularly interesting in the context of this study as in all cases it is in the extended wings that we measure offsets in position with respect to the source. 
}
\maketitle
\section{Introduction}
The phenomena of Herbig-Haro (HH) jets and molecular outflows from Young Stellar Objects (YSOs) have been much studied since the 1980's and in recent years great progress has been made towards increasing our understanding of the mechanisms behind them. Models of jet/outflow acceleration, collimation and evolution are constrained through observations of these phenomena close to their driving sources i.e. to within a few hundred AU. The technique of long-slit spectroscopy is a particularly useful tool in this work as it allows us to examine the structure and kinematics of the jets and outflows on the smallest spatial scales available, while minimising any contrast problem with the source.

 Numerous studies have shown, for example \cite{Hirth94(a)}, \cite{Hirth94(b)} and \cite{Davis03}(hereafter DWRC), that Forbidden Emission Lines (FELs) serve as powerful tracers of outflow activity close to the central engines of YSO's. Thus they are ideal for use with the long-slit spectroscopic technique. One important characteristic of FEL profiles is that they are often multi-component. A high velocity component or HVC and a low velocity component or LVC are seen. For example in a study of 38 T-Tauri stars completed by \cite{Hirth97}, 50$\%$ of the T-Tauri stars in a sample of 33 objects with high S/N [OI]$\lambda$6300 line profiles, showed a double peaked profile. Popular consensus holds with the theory that the HVC represents a high velocity well-collimated jet and the LVC a broad disk wind \citep{Kwan88, Kwan97}.

In the case of the T-Tauri stars atomic hydrogen emission lines are also useful as a means of probing activity close to the central engines. Strong hydrogen emission lines are a hallmark feature of Classical T-Tauri stars (CTTSs) and their profiles are strongly influenced by the dynamics of the regions in which they form. In contrast to FELs however there is no consensus on how or where they are excited. Explanations based on their formation in outflows \citep{Natta88, Hartmann90, Calvet92} and in accretion zones as described by magnetospheric accretion model (MAM) \citep{Calvet92, HHC94, Muzerolle98(a)} have both been suggested. It seems likely that no model based solely on a mass loss or mass accretion scenario can account for all the different HI emission lines detected. 

To date most work has been done using the Balmer lines, especially H$\alpha$ as a reference. Although H$\alpha$ clearly traces outflows at large distances from the source, close to the source the situation may be different \citep{Calvet97}. While P-Cygni H$\alpha$ profiles are clearly formed by winds it is also obvious from studying Inverse P-Cygni profiles that the H$\alpha$ line can be shaped by accretion. For profiles that are not P-Cygni in shape however questions still remain as to their origin. Recent studies have clearly shown that MAM cannot entirely explain HI emission lines from for example the Brackett or Paschen series of lines also. An important study by \cite{Folha01}, looked at the Pa$\beta$ line profiles of 49 T-Tauri stars and highlighted the differences found between predictions made by models and observations. They concluded that Pa$\beta$ emission may partly trace the outflow and that current models do not go far enough in accounting for the formation and characteristics of these lines. We will  discuss work of \cite{Folha01} in more depth in section 4.2. \cite{Alencar01} also conclude from their study of permitted emission from T-Tauri stars that the MAM is only part of the process that produces strong permitted line emission in CTTSs and that part of the emission line profiles may also be produced in winds. 

The main approach of this work is to compare results obtained from a study of a small sample of TTSs in Pa$\beta$ with optical FEL observations taken of the same sources (in some cases [FeII](1.644$\mu$m) data is also used as a comparison). The aim is to see if the Pa$\beta$ line traces the outflow from these YSOs as effectively as the optical FELs. If Pa$\beta$ can be used to study outflow activity from T-Tauri stars on small spatial scales then we have conclusive evidence to suggest that an explanation of the formation of HI lines based entirely on the MAM is flawed. Also this will make Pa$\beta$ a very useful line for use in conjunction with the new class of near-infrared interferometers, for example AMBER, now coming online \citep{richichi00}. They will give us very high spatial resolutions and as our results suggests Pa$\beta$ traces the outflows close to their source, it could be a very useful probe of the central engine. 

\begin{table} 
\begin{tabular}{llll}
\hline \hline                                               
 Source           &RA (J2000)    &Dec (J2000)       &Slit P.A.                                                               
\\ \hline 
 DG Tau           &04 27 04.7    &+26 06 17         &$46^{\circ}$
\\
 V536 Aql        &19 38 57      &+10 30 16.6       &$110^{\circ}$
\\
 LkH$\alpha$ 321 &21 01 55.1    &+49 51 35         &$208^{\circ}$
\\
 RW Aur          &05 07 49.6    &+30 24 05         &$125^{\circ}$
\\
 \hline               
\end{tabular}
\caption{Right Ascension, Declination and Position Angle (PA) of all sources included in this study.}
\end{table}

\section{Observations/Data Reduction}
Near-infrared Pa$\beta$ observations were acquired for the four T-Tauri stars listed in Table 1. These observations were complemented by optical forbidden emission line data. However only 3 out of 4 of the sources were observed in the optical, namely DG Tau, V536 Aql and LkH$\alpha$ 321. Observations taken during three different runs make up our data set. 

Firstly the Pa$\beta$ observations of DG Tau and RW Aur were obtained at the United Kingdom Infrared Telescope (UKIRT) on 3 November 2001  using the CGS4 spectrometer. The [FeII] data presented as a comparison for both these sources was also taken on this night. CGS4 is equipped with a 256 $\times$ 256 InSb array. The pixel scale at 1.2822$\mu$m and 1.644$\mu$m measures 0$\farcs$88 in the spatial direction. The slit width was two pixels giving a velocity resolution of $\sim$ 7.5 km/s and $\sim$ 8km/s per pixel for Pa$\beta$ and [FeII] respectively. The instrumental profile in the dispersion direction measured from gaussian fits to arc lines was 19.0($\pm$1.0 km/s). In each observation the slit was orientated so that the source was at the centre of the slit and along the jet axis. Position angles for each of the sources are shown in Table 1. Secondly the Pa$\beta$ observations of V536 Aql and LkH$\alpha$ 321 were made on 7 August 2002  during a service run on UKIRT. Again the CGS4 spectrometer was used and the slit width, pixel scale and dispersion were the same as for the November run. 

In both cases a sequence comprising one sky followed by three object exposures was repeated a number of times for each source, in order to build up signal to  noise. The sky position was typically 30$\arcsec$-60$\arcsec$ away from each source in a direction orthogonal to the flow axis. Each image was bias subtracted, flatfielded and co-added into reduced groups. The data reduction was done using standard IRAF routines. In the case of the Pa$\beta$ observations each source was wavelength calibrated using an Krypton Argon arc that was obtained immediatley prior to observing each spectrum. Only two arc lines, spread across the dispersion axis, could be used for wavelength calibration. They were the krypton line at $\lambda_{vac}$ = 1.27859$\mu$m and the argon line at $\lambda_{vac}$ = 1.28062$\mu$m. The [FeII] data was calibrated as described in DWRC. 

The optical FEL data was taken during service runs on the  Newton Telescope (INT) using the Intermediate Dispersion Telescope (IDS). Both DG Tau and LkH$\alpha$ 321 were observed on 17 September 2002 and V536 Aql on 10 October 2002. With the IDS we used the 500mm camera, the R1200Y grating and the EEV10 CCD chip. This gave a central wavelength of 6500\AA, a spatial resolution of 0$\farcs$19 and an approximate dispersion of 0.2214\AA   per pixel. Again in order to reduce the grouped, bias subtracted and flat-fielded images we used standard IRAF routines. The spectra from DG Tau and LkH$\alpha$ 321 were calibrated using a Copper-Argon arc spectrum. V536 Aql was calibrated using a Copper-Neon arc. 

The absolute velocity calibration for all sources was checked by self-calibrating the arc spectrum. For the Pa$\beta$ data, variations in the calibration were found to be small, only of the order of 3km/s. As reported in DWRC the corresponding values for the [FeII] data were of the order of 5km/s. In the case of the optical data the CuAr and CuNe arcs gave a variation of less than 1km/s. Variations were much less towards the centre of the array. 

\begin{figure*}
\epsfig{file=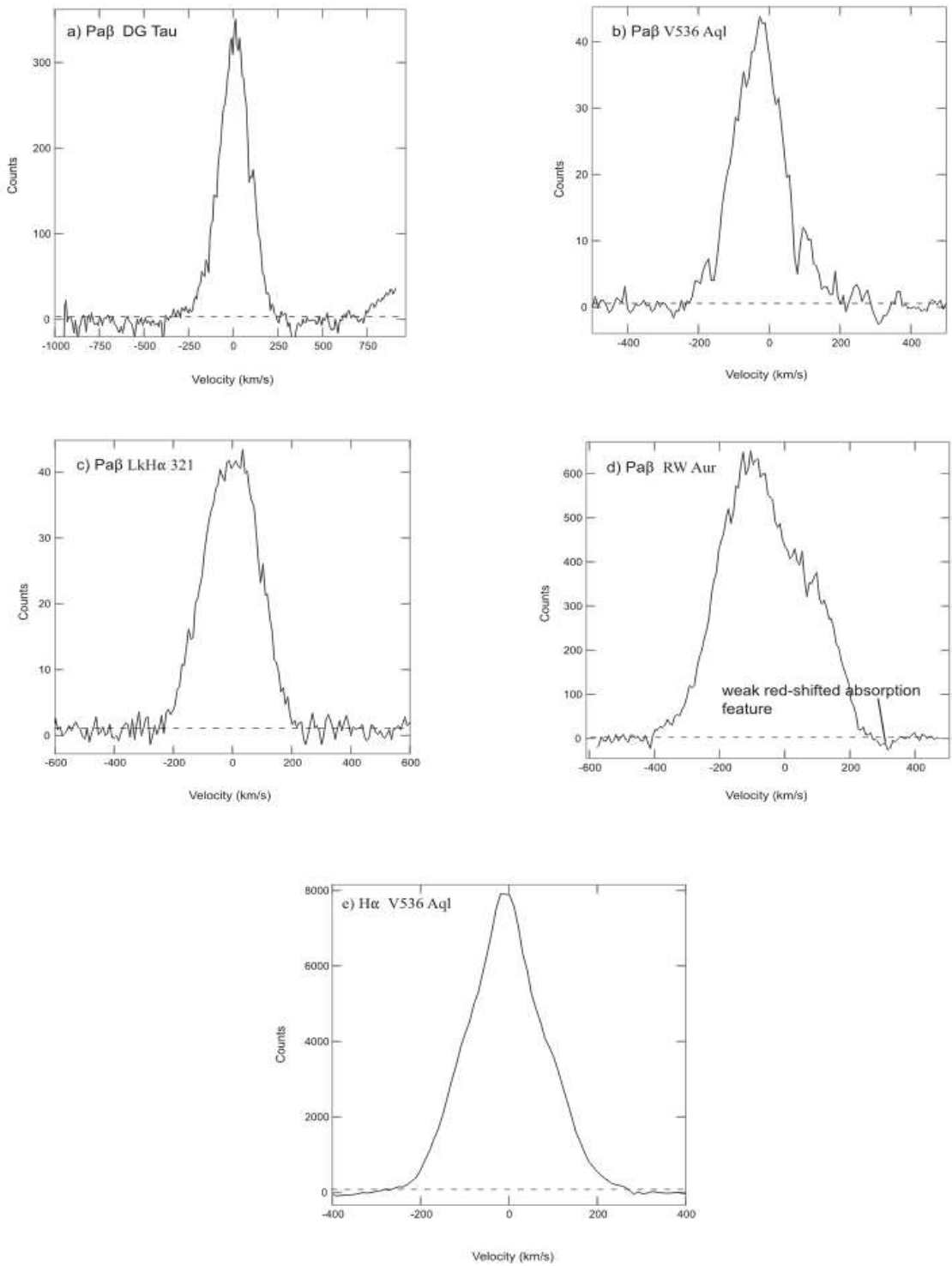,width=17cm,height=20cm}
\caption{Continuum subtracted Pa$\beta$ emission line profiles for the 4 T-Tauri stars and the continuum subtracted H$\alpha$ line profile for V536 Aql. The position of the 3-sigma background noise level is marked by the dashed line.}
\end{figure*}

\begin{figure*}
\epsfig{file=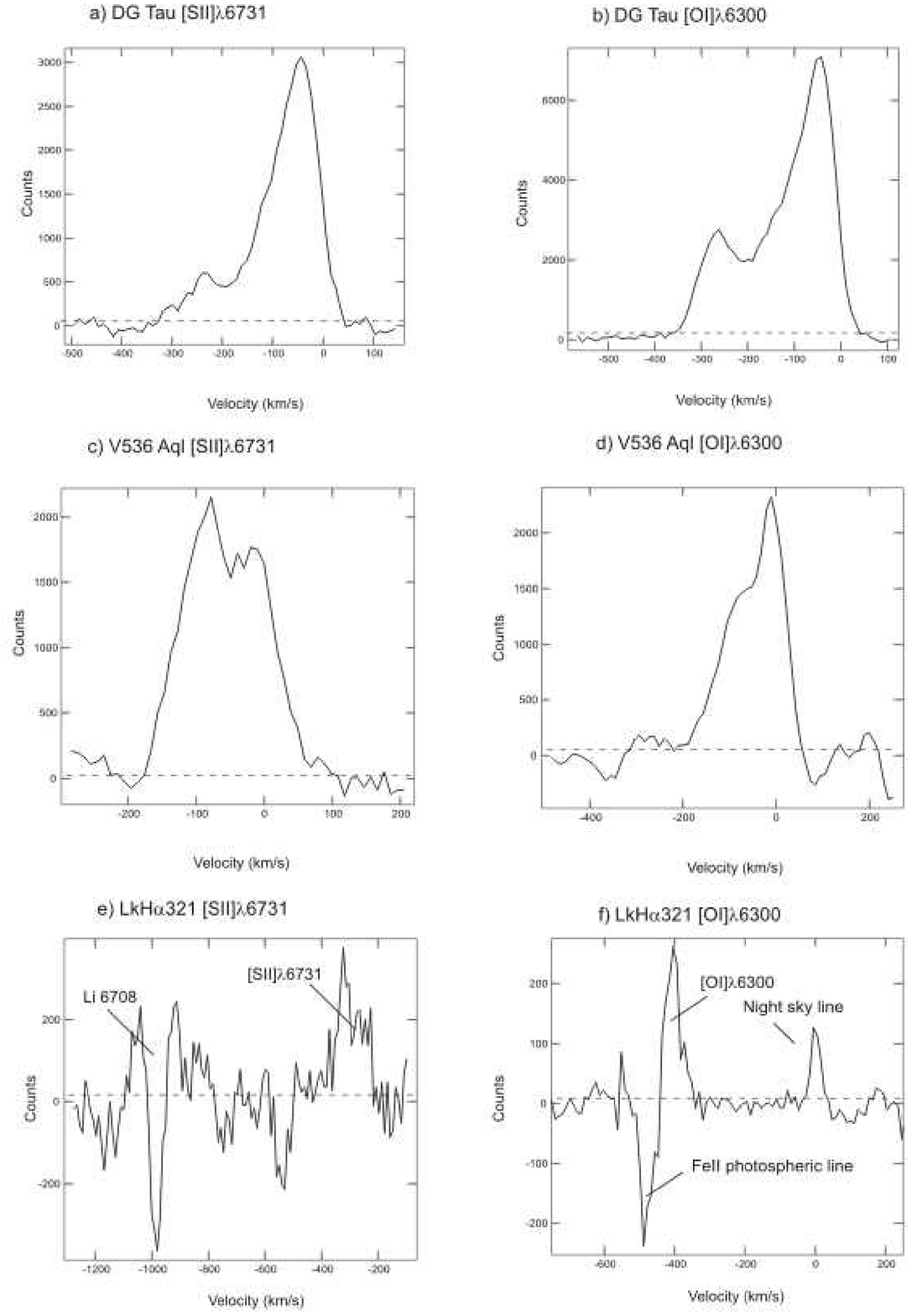,width=15cm,height=20.5cm}
\caption{Continuum subtracted velocity profiles of the optical FELs from DG Tau, V536 Aql and LkH$\alpha$ 321. Line profiles represent 1-D spectra extracted close to the source and are the sum of three rows. The position of the 3-sigma background noise level is marked by the dashed line.}
\end{figure*}

\section{Results}
As stated above the main aim of this study is to directly compare optical FEL spectra (and in some cases [FeII] spectra) with Pa$\beta$ spectra. With this in mind we will present our main results in the form of spectro-astrometric plots, showing the relative position of the FEL and Pa$\beta$  emission with respect to the source continuum position. In order to measure the positional displacement of the emission we first take an accurate measure of the centre of the continuum using gaussian fits and set this to zero offset. We then subtract the continuum from the spectrum (as the presence of the continuum will tend to drag the position of the emission back towards the source) and extract one column. Each column represents a particular velocity (in some cases where emission was particularly weak it was necessary to sum over 3-6 columns) and we find the position of the emission with respect to the centre of the continuum at this velocity by fitting a gaussian to this column. In this way we record the distance of the emission centroid from the centre of the continuum over a velocity range. The error in the measured displacement of the emission centroid is given by seeing/$(2(2ln2)^{1/2}\sqrt{N})$, where N is the number of detected photons. Using this method we measure an average error of 0$\farcs01$. 

We shall also present figures showing the Pa$\beta$ and FEL line profiles. Continuum-subtracted position velocity diagrams were prepared for all the optical FELs that were found to be bright close to the sources although only a limited number are presented here. Figure 1 and 2 present the Pa$\beta$ and optical FEL profiles of the sources listed in Table 1. Figure 3 and 4 show the continuum-subtracted position-velocity diagrams of the optical FELs from DG Tau and V536 Aql. The results of applying the spectro-astrometric technique to the four CTTSs are diaplayed in Figure 5 and 6. All velocities quoted here are local standard of rest (LSR) velocities. We estimated the systemic LSR velocity of V536 Aql and LkH$\alpha$321 using the photospheric Li$\lambda$6708 absorption line.

\subsection{DG Tau (HH 158)}
DG Tau is one of the closest classical T Tauri stars at a distance of $\sim$140 pc and it has been shown to drive a well-collimated HH jet (HH 158). It was amongst one of the first CTTSs from which a jet like outflow was discovered \citep{Mundt83}. The jet which has been well studied spectroscopically at optical forbidden emission lines, firstly at medium resolution \citep{ Mundt87} and later at high spatial and spectral resolution \citep{Lavalley97, Eisloffel98} extends to approximately 12 \arcsec  in a southwest direction. \cite{Eisloffel98} detected four knot like features labelled A,B,C,D and suggested that the jet terminates in a bow shock at Knot C. Knot D was described as a faint feature close to the apex of the bow shock associated with Knot C and because of its location it was believed that it resulted from some irregularities. \cite{Lavalley97} presented further data on two knots close to the star i.e. Knot A and B mentioned above and concluded that Knot B also had a velocity gradient and morphology consistent with it being a bow shock. \cite{Solf93} studied the optical forbidden emission from the jet close to the source i.e within 0\farcs5 and found evidence for two separate gas components in the flow, a low and a high velocity component, following the model of \cite{Kwan88}. Lastly, further studies have been carried out with HST by \cite{Bacciotti00}, \cite{Bacciotti02}, showing the jet structure to within a few tenths of an arcsecond of the source and pointing to possible rotation in the jet.

DG Tau was also one of the T-Tauri stars included in our [FeII] (1.644 $\mu$m) study of T-Tauri stars (DWRC). We detected blue-shifted [FeII] emission predominantly from the DG Tau jet. The wavelength coverage of these echelle observations included the Pa$\beta$ near-infrared line and  here we report the detection of Pa$\beta$ emission from DG Tau (figure 1(a)). It has a FWHM of $\sim$ 225 km/s and a blue-wing that extends to approximately -380 km/s. It is centred at $V_{LSR}$ $\sim$8 km/s. For DG Tau the systemic LSR velocity is 6 km/s \citep{Kitamura96}. Figure 5(a)(i) shows the result of applying the spectro-astrometric technique to the Pa$\beta$ emission from DG Tau. We can clearly see that we measure offsets in the blue-shifted wing out to about 0\farcs5 from the source and that the offset peaks at a velocity of$\sim$-250 km/s.

We also present the results of our recent optical long-slit spectroscopic observations of this YSO. We detect [SII]$\lambda\lambda$6716, 6731, [OI]$\lambda\lambda$6300, 6363, [NII]$\lambda\lambda$6548, 6583 and H$\alpha$ emission from the jet. We also detect all the above lines except the [NII]$\lambda$6548 line towards the source. At the source, in [SII]$\lambda$6731 we see a low velocity component at -47 ($\pm$ 7) km/s and a high velocity peak at -240 ($\pm$ 6) km/s. See figure 2(a). The [SII]$\lambda$6716 emission line has a similar profile, with a LVC at -42 ($\pm$ 7)km/s and an extended blue wing. The [OI]$\lambda$6300 is double peaked. The low velocity component has a centroid velocity of $\sim$-47 km/s ($\pm$ 6) km/s and the high velocity component peaks at $\sim$-266 ($\pm$ 6)km/s. The [OI]$\lambda$6363 line peaks at $\sim$ -55 ($\pm$ 6) km/s and has a blue wing that extends out past -300 km/s. Lastly the [NII]$\lambda$6583 line traces high velocity emission only, peaking at a velocity of $\sim$-257 ($\pm$ 10) km/s. Figure 3 presents the continuum subtracted position velocity diagrams for the [SII]$\lambda$6731 and [OI]$\lambda$6300 lines. We can see Knots B and C mentioned above. See Figure 5(a)(iii) and (iv) for the spectro-astrometric plots of the [SII]$\lambda$6731 and [OI]$\lambda$6300 lines.   
\subsection{V536 Aql}
Early studies of the Classical T-Tauri star V536 Aql showed it to have relatively strong forbidden line emission in [OI]$\lambda$6300 \citep{Cohen79} and a large degree of polarisation \citep{Bastien82}. This strong degree of polarisation was suggestive of the presence material distributed around the star. \cite{Ageorges94} presented the first evidence for an extended circumstellar medium around V536 Aql. They also presented in their paper high angular resolution images that showed for the first time that V536 Aql is a pre-main sequence binary. The two stars are separated by 0$\farcs$52 at a PA of $17^{\circ}$. The spectroscopic study of \cite{Hirth97} provided the first conclusive evidence of the presence of a bipolar outflow from V536 Aql. They studied it in various optical FELs and showed that it is spatially extended by about 3\arcsec-4\arcsec  at a PA of $90^{\circ}$ $\pm$ $20^{\circ}$ for the blue-shifted part of the flow. Lastly the [SII]$\lambda\lambda$6716, 6731 and continuum images taken by \cite{Mundt98} confirmed the findings of \cite{Hirth97}. The images showed several knots at a position angle of $110^{\circ}$. The nearest of these knots was at about 4$\farcs$2 and the furthest was located at about 16$\arcsec$. In the [SII] image a faint counterflow was seen.

V536 Aql was not one of the sources that was part of our original [FeII] study instead it was included in our service night observations on UKIRT. We detect no [FeII] emission from V536 Aql but we did detect Pa$\beta$. See Figure 1(b) for the Pa$\beta$ line profile. It has a FWHM of $\sim$167 km/s centred at $\sim$-30 ($\pm$ 9)km/s and it has a red wing that extends to $\sim$300km/s. When we applied the spectro-astrometric technique to this source we measured offsets for both the red and blue lobes of the flow. See Figure 5(b)(i). We estimate the systemic LSR velocity of this source at approximately -2 km/s.   

Our FEL observations of this source in the optical gave the following results. We detected  [SII]$\lambda\lambda$6716, 6731, [NII]$\lambda\lambda$ 6548, 6583, [OI]$\lambda\lambda$6300, 6363 emission from the extended outflow but only out to about 3$\arcsec$. The [SII]$\lambda$6731 and [OI]$\lambda$6300 lines made up the strongest emission from close to the source. In the case of [SII]$\lambda$6731 the line profile is double peaked with a HVC at -84 ($\pm$ 5) km/s and a LVC at -20($\pm$ 5) km/s. See Figure 2(c). Hirth et al 1997 report the [OI]$\lambda$6300 emission line as being double peaked with a LVC at -7 km/s and a HVC at -77 km/s (data was taken in November 1993). We detect a strong LVC at $\sim$-10 ($\pm$ 7)km/s and a blended HVC which we estimate to be at $\sim$-65 km/s. See Figure 2(d). We again prepared position velocity diagrams for the [SII]$\lambda$6731 and [OI]$\lambda$6300 emission (see Figures 4(a) and (b)). See Figures 5(b)(iii) and (iv) for the plots of offset of the [SII] and [OI] emission against velocity. 

Interestingly the H$\alpha$ emission from this star has a similar profile to the Pa$\beta$ profile. It has FWHM of 200 km/s, is centred at $\sim$-7 ($\pm$ 9)km/s and it has symmetric red and blue wings that extend to approximately -300km/s. Again spectro-astrometry reveals a displacement in both the red and blue wings of this emission. See Figure 5(b)(ii).

\subsection{LkH$\alpha$ 321}
This YSO has been little studied to date. \cite{Mundt98} classify it as a CTTS however \citep{Herbst99} decscribe it as being intermediate between a lower mass and higher mass YSO. Indeed the luminosity and mass of this source (146L$\sun$, 3M$\sun$) \cite{Chav81} is consistent with a higher mass source. The distance of its parent cloud has been estimated at $\sim$ 550pc \citep{Shev91}. It was shown by \cite{Cohen79} to have FELs of only moderate strength. \cite{Mundt83} revealed that it had prominent blue-shifted absorption features in the Na I$\lambda\lambda$5890, 5896 absorption lines. \cite{Mundt98} published [SII]$\lambda\lambda$6716, 6731 and continuum images. The [SII] and the [SII]-continuum images showed a knotty jet extending 22$\arcsec$ to a very faint knot labelled knot E at a PA of $208^{\circ}$. The only evidence for a counterflow is a faint knot labelled knot F seen only in the [SII]-continuum image. We estimate the systemic velocity of this source to be $\sim$9 km/s

Again the near-infrared observations of this source were taken on UKIRT using CGS4 during a service run. We did see this source in [FeII] however only towards the source and not from the extended outflow. The [FeII] emission from LkH$\alpha$321 is red-shifted and peaks at a LSR velocity of 229 ($\pm$ 4) km/s. We do not see the blue-shifted flow in [FeII]. We conclude that there is either no blue-shifted emission or else it is blended with the strong Brackett 12 line (1.641168$\mu$m) which we also detect. The red-shifted [FeII] was far too weak with respect to the continuum to be able to measure any positional displacement in it. In Pa$\beta$ we again see a broad profile with a FWHM of approximately 300 km/s that is centred on $\sim$-3 ($\pm$ 10) km/s. See figure 1(d). Spectro-astrometry reveals offsets in emission tracing both red and blue-shifted outflowing material. In the blue-wing offsets peak at approximately 160km/s. The corresponding value for the red-wing is 120 km/s. 

Again our optical observations of the source were made with the IDS on the INT. We detect [SII]$\lambda\lambda$6716, 6731, [NII]$\lambda$6583, H$\alpha$ and [OI]$\lambda$6300 emission. The [SII]$\lambda$6731 and the [OI]$\lambda$6300 lines are the strongest FELs close to the source but in comparison to what we see in DG Tau or V536 Aql the emission is very weak (see Figure 2(c) and 2(d)). Hence it was difficult to accurately measure offsets in these lines especially in the case of [SII]$\lambda$6731. However we can still see a clear offset. Spectro-astrometric plots of these lines are presented in figure 6. From 6(a) (ii) the [SII]$\lambda$6731 line peaks at a velocity of -340 km/s. We estimate the peak LSR velocity of the [OI]$\lambda$6300 line to be at -412($\pm$ 5) km/s, by fitting a gaussian to its profile (see also Figure 6(a)(iii)). This is possible as it is much stronger close to the source than the [SII]$\lambda$6731 line as is apparent from Figure 2.

\subsection{RW Aur (HH 229)}
RW Aur is located in the Taurus-Auriga system at a distance d$\sim$ 140pc. It is a complex multiple star system with the primary or brightest stellar component A  having a separation of about 1$\farcs$4 from the close binary B$\&$C (separation 0$\farcs$12) \citep{Ghez93}. \cite{Ghez93} made their observations in the K-band and components were assigned the letters A, B, C based on the relative K-band flux of the stars at the time, with A being the brightest. Component A may itself be a spectroscopic binary \citep{Gahm99}. The HH229 jet is a bipolar outflow associated with component A. It was first discovered by \cite{Hirth94(b)} by long-slit spectroscopy in H$\alpha$ and several forbidden emission lines. Their results showed that the southeastern blue-shifted jet extended over 22$\arcsec$. Later studies by \cite{Eisloffel98} however suggest a total  length of at least 145$\arcsec$ for the bipolar outflow. A new study by McGroarty et al 2003 (in preparation) puts this length at approximately 7$\arcmin$.

RW Aur formed part of our original 2001 [FeII] study (DWRC) and [FeII] results presented here were already published in this paper. In [FeII] we detected emission from both lobes of the jet. Figure 6(b)(ii) shows the plot of the offset of the emission with respect to the source continuum position. In the red-shifted lobe we measure a peak radial LSR velocity of 150 ($\pm$ 10) km/s extending to almost 0$\farcs$8. The systemic LSR velocity for this source is approximately +6 km/s \citep{Ungerechts87}. In the blue-shifted lobe the LSR radial velocity peaks at approximately -175 ($\pm$ 10) km/s and reaches an offset of $\sim$1$\farcs$2. 

We also detected Pa$\beta$ emission from RW Aur. The line profile is shown in Figure 1(d). \cite{Folha01} classified this line profile as an Inverse P Cygni profile (see section 4.2), however its red-shifted absorption feature is very weak. It has a FWHM of $\sim$320 km/s and its blue-wing extends to $\sim$-400 km/s. When we used the spectro-astrometric technique on this source we found that we could measure no offsets in the Pa$\beta$ emission. See Figure 6(b)(i).
\begin{figure*}
\epsfig{file=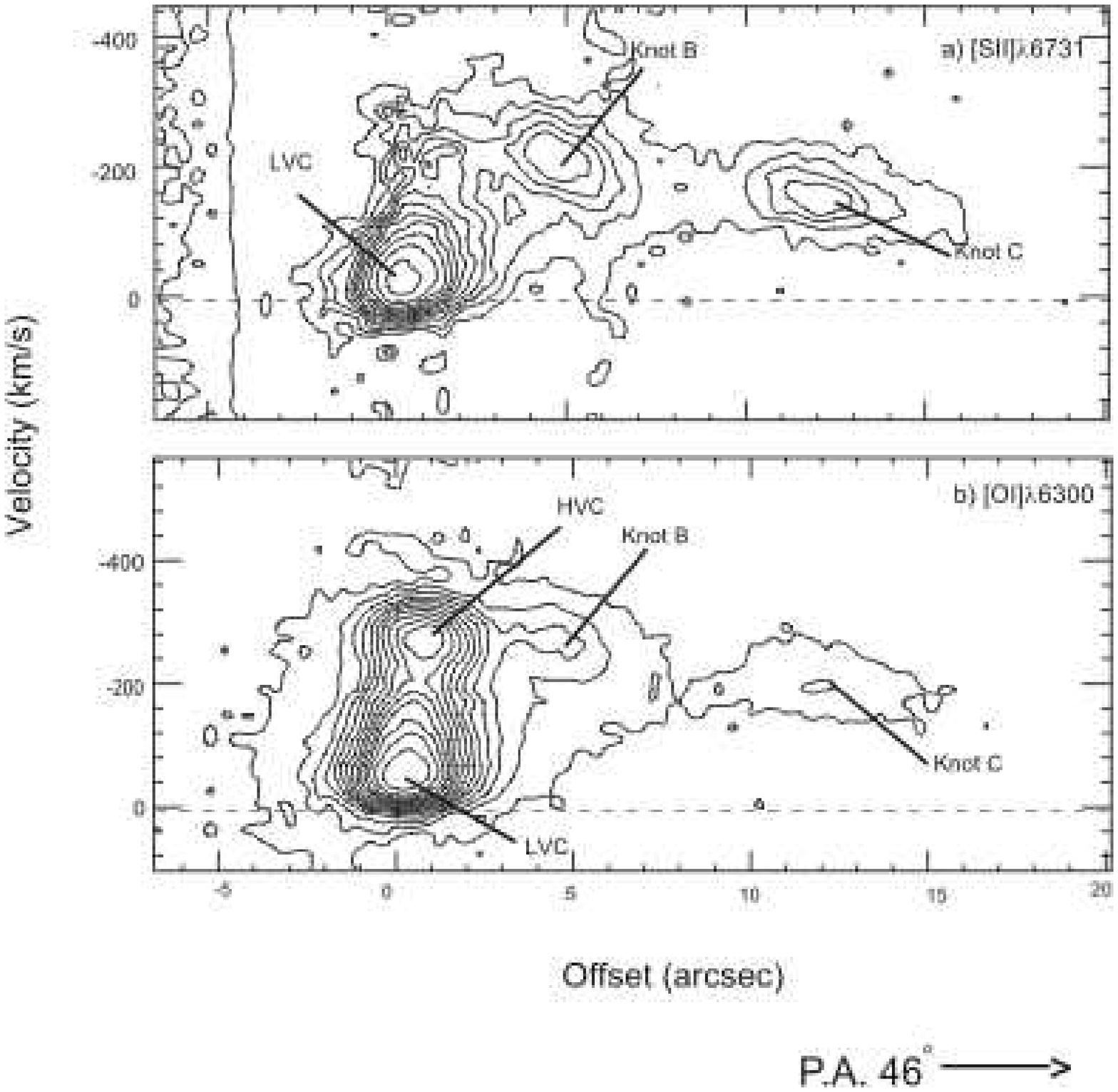,width=13cm,height=10cm}
\caption{Continuum-subtracted position velocity diagrams of optical forbidden emission lines from DG Tau. Contours start at 3 times the standard deviation of the background noise and increase by a factor of $\sqrt{2}$. The dashed lines mark the systemic LSR velocity of the source.}
\end{figure*}

\begin{figure*}
\epsfig{file=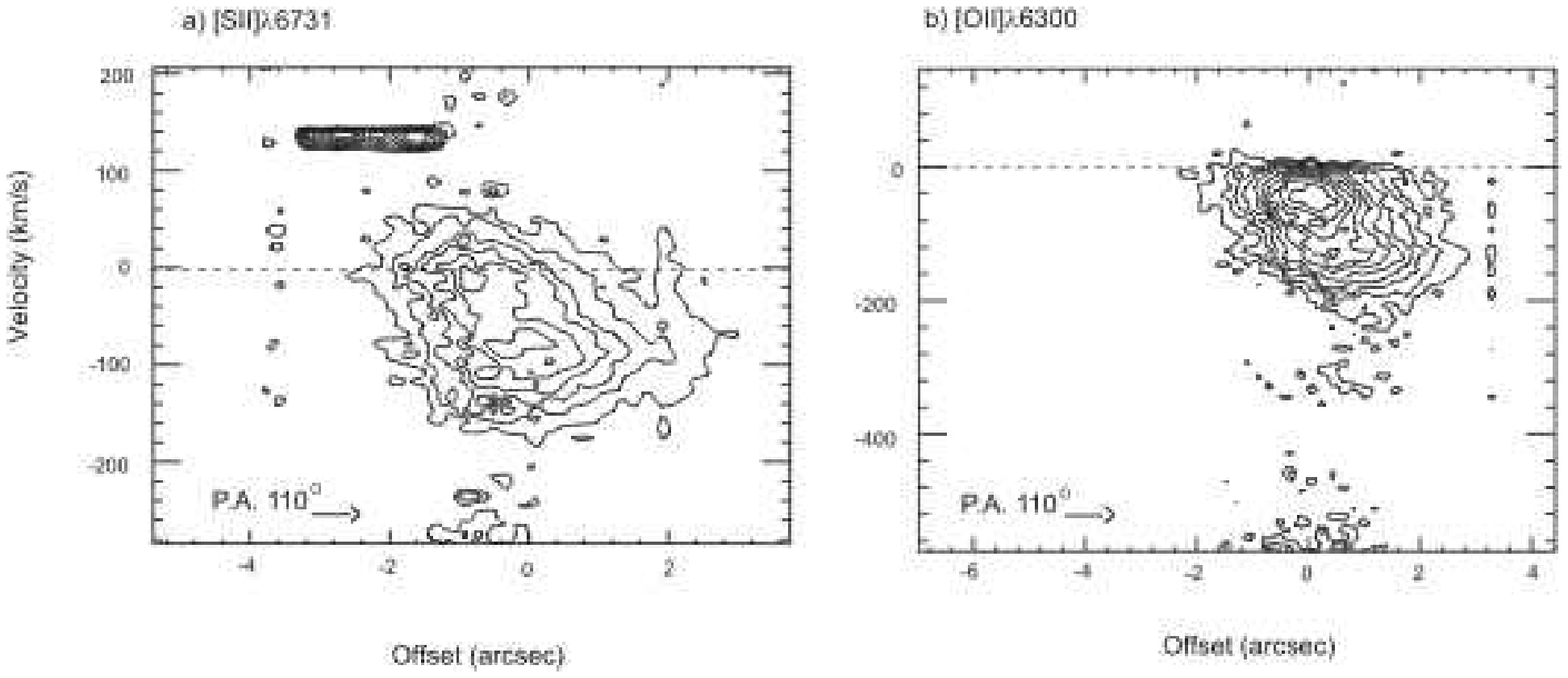, width=17cm,height=8cm}
\caption{Continuum-subtracted position velocity diagrams of the optical forbidden emission lines from V536 Aql. Contours start at 3 times the standard deviation of the background noise and increase by a factor of $\sqrt{2}$. The dashed lines mark the systemic LSR velocity of the source.}
\end{figure*}

\begin{figure*}
\epsfig{file=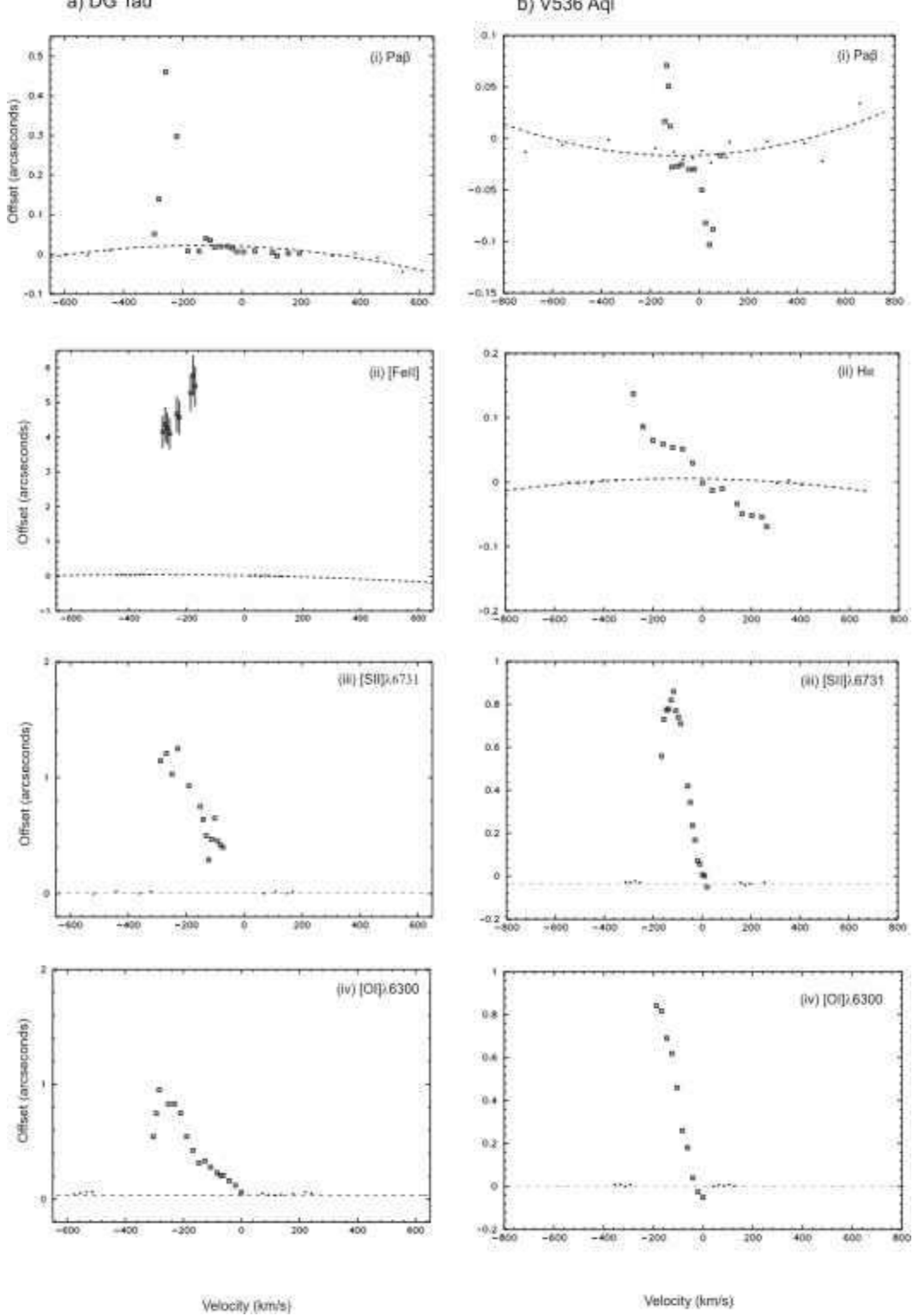, width=14.cm,height=20.5cm}
\caption{Spectro-astrometric plots of Pa$\beta$, [FeII] 1.644$\mu$m, [SII]$\lambda$6731 and [OI]$\lambda$
6300 emission from DG Tau and V536 Aql. The PAs are 46 and 110 degrees for DG Tau and V536 Aql, in the positive offset axis direction, respectively. All velocities are LSR velocities. The systemic velocity of each source is quoted in the results section. The uncertainity in position is typically 0$\farcs01$ as described in the results section, however for weaker emission for example Figure (iii) this value is obviously higher. Both sources have low velocity optical forbidden emission as is clear from their line profiles in Figure 2, however the emission is blended giving a continuous increase in offset as seen in the above plots. The centroid velocity of the low velocity emission was measured from the emission line profiles and using the above plots we could estimate the position of this emission. See Table 2.}
\end{figure*}

\begin{figure*}
\epsfig{file=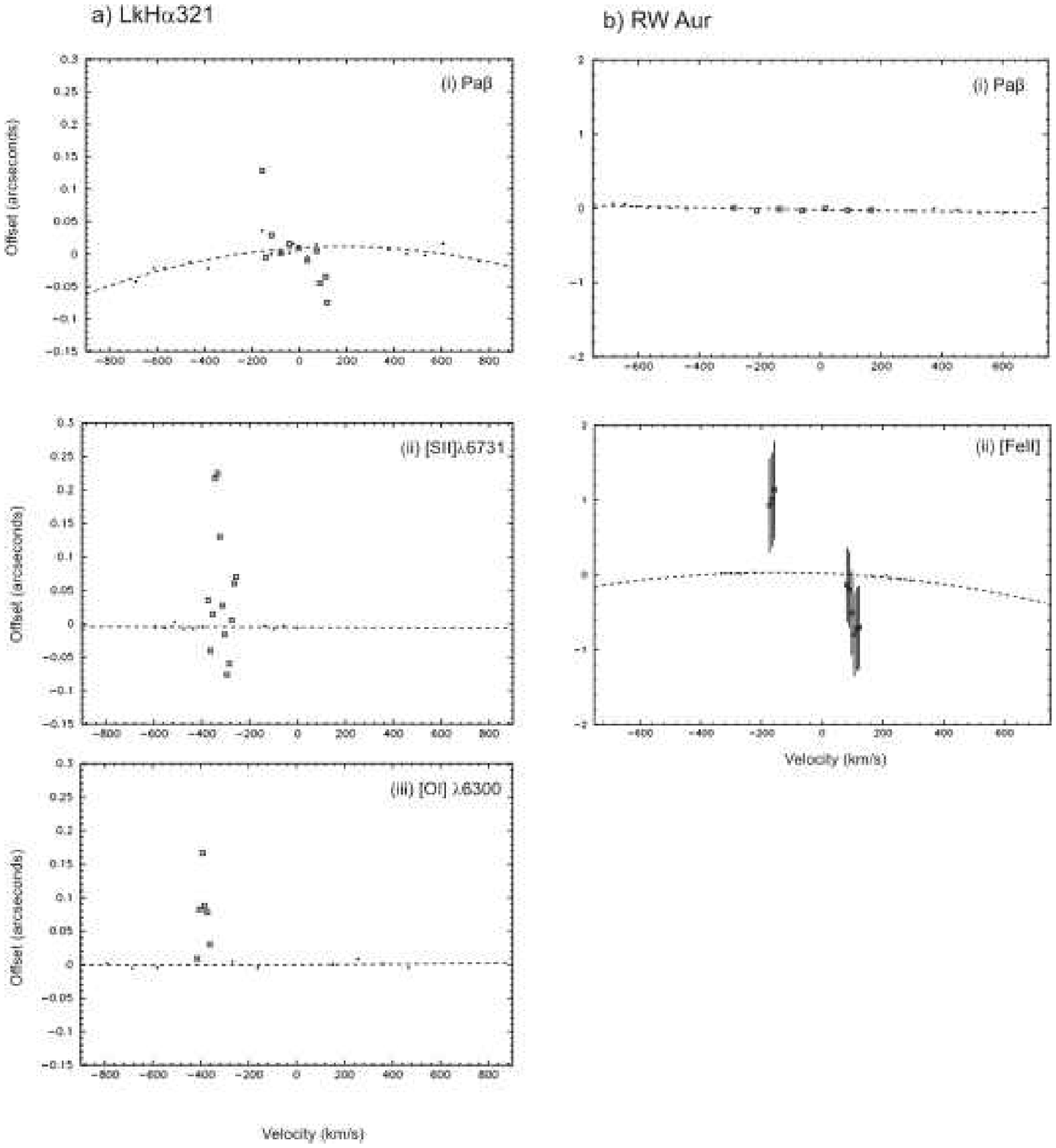,width=14cm,height=15.25cm}
\caption{Spectro-astrometric plots of emission from LkH$\alpha$321 and RW Aur. The PAs are 208 and 125 degrees, in the direction of the positive offset axis direction, respectively. All velocities are LSR velocities. The systemic velocity of each source is quoted in the results section.  The uncertainity in position is typically 0$\farcs01$ as described in the results section. }
\end{figure*} 

\section{Discussion}
\subsection{Explaining HI emission lines}
Considerable effort has been made over the last 20 to 30 years to try and model atomic hydrogen emission lines from T-Tauri stars. To date the principal models have involved magnetically driven winds and more recently magnetospheric accretion. A good understanding of how these HI emission lines form is important as it enables us to use them as diagnostic tools. 

The Balmer series of lines and in particular the H$\alpha$ line have been the basis of the majority of studies to date. The major advantage of the H$\alpha$ line for high resolution work is that it is generally the brightest spectroscopic feature seen in YSOs and the most common. Its wavelength falls near the peak sensitivity of a CCD hence one can observe it even when it is relatively faint in absolute terms. Historically it was easier from a technical point of view to observe emission lines in the optical, as techniques for observing emission in the NIR, for example, were not well developed.

\cite{Reipurth96} presented an extensive study of H$\alpha$ emission from YSOs. They categorised the profiles into four different types in an aim to  propose a scheme that reflected the underlying physical processes responsible for the line, yet was also able to deal with a greater variety of profiles. Type I profiles are normally symmetric, slightly blue-shifted (they can however be occasionally red shifted) and show no absorption features. Type II are double peaked with the intensity of the second peak exceeding half the strength of the main peak. Type III profiles are also double peaked but the intensity of the second peak is less than half the intensity of the main peak. Types II are III assigned the letters R or B depending on whether the secondary peak is blue or red-shifted with respect to the primary peak. Lastly Type IV profiles are similar to Type I but they show an absorption feature which marks the edge of the emission. In other words depending on whether this absorption feature is blue or red-shifted the Type IV profile corresponds to either a P Cygni (PC) or Inverse P Cygni (IPC) profile and is appended with either the letter B or R. Their results showed that 25$\%$ of T-Tauri stars had a Type I profile and 33$\%$ had a Type IIIB profile. They found that very few CTTSs in their sample had a PC or IPC H$\alpha$ line profile. 

\cite{Reipurth96} also discussed in depth the different models used over the years to explain HI (Balmer) emission lines. We  will now briefly summarise some of these models in order to give an idea of the history of the debate on the formation of HI lines. In general IPC and PC profiles are easily explained through accretion and wind models respectively. The features of the lines that models have struggled to explain are their large Full Width Half Maximum (FWHM), their symmetry and the presence of extensive emission wings. Type I profiles incorporate all these features.

The majority of mass loss models assume the wind is magnetically driven. Thermally driven winds were discounted as they could not reproduce the broad line-widths seen. Early models suggested Alfv\'{e}n waves propagating outwards as the wind driving mechanism \citep{Lago79}, \citep{DeCampli81}. These models did not attempt to fit the H$\alpha$ line as it is optically thick thus an adequate treatment of radiative transfer had to be included. Higher Balmer lines that were assumed to be optically thin were used instead.  

Improvements in the treatment of radiative transfer were made by \cite{Natta88} and \cite{Hartmann90}. \cite{Hartmann90} used a spherical wind model where again the driving mechanism was Alfv\'{e}n waves. Overall, models did not match the observations well as the computed H$\alpha$ and H$\beta$ profiles were broader and less symmetric than observed with blue-shifted absorption features that were also broader and deeper than seen.  

\cite{CalvetH92} described a wind, that originated in a boundary layer between a star which is rotating slowly and a disk which is rapidly rotating and flowed occupying a cone-like  geometry. Models produced central and blue-shifted absorption but the level of absorption exceeded that observed. Also centrally peaked Type I profiles could not be reproduced. They tried to do this by including a rotational velocity component but this failed. 

\cite{Calvet92} suggested a different approach based on the MAM, see also \cite{Koenigl91} and \cite{Shu94}. They suggested that accretion is restricted to an axially symmetric configuration by the dipolar stellar magnetic field. This magnetic field truncates the disk at a certain radius and hence there are magnetic field lines connecting the star to the disk. Accretion onto the star occurs along these magnetic field lines. These models were improved upon by \cite{HHC94} and \cite{Muzerolle98(a)}. Muzerolle et al (1998) improved further on the work by considering a multi-level atom in equilibrium rather than the 2-level model used by the earlier authors. It was also noted that the size of the inner and outer radii of the magnetosphere, the temperature and the inclination angle all influence the profile of the Balmer lines. A good overall agreement was found between observations and computed profiles especially in the case of IPC profiles. The major discrepancy lay in the fact that they could not reproduce the line wings. They suggested that the high velocity wings were not actually due to the higher velocity infalling material but due to Stark broadening. This hypothesis was tested but it was actually found that the contribution was small. They also suggested that the emission that makes up the line wings may actually result from dynamic zones other than zones of infall i.e. zones of outflow.

In summary infall models have produced the most promising results but discrepancies such as large FWHMs and the presence of line wings still persist. Even though IPC profiles can be naturally explained by infall models Type I profiles cannot be accounted for. Recently work has been extended to other line series for example the Brackett series \citep{Najita96} and more importantly in the context of this paper the Paschen series \citep{Folha01} which we shall discuss next.   

\subsection{Folha and Emerson study}
Folha and Emerson (2001) carried out the first extensive study of Pa$\beta$ and Br$\gamma$ near-infrared hydrogen emission lines from T-Tauri stars. We will concentrate on their findings for the Pa$\beta$ line. They firstly sorted the emission line profiles according to the classification scheme proposed by \cite{Reipurth96}, described above. Pa$\beta$ emission from 49 CTTSs was studied. Of the 49 stars the authors concluded that 53$\%$ of the Pa$\beta$ line profiles were Type I and 34$\%$ were Type IVR (IPC) profiles. Overall the Type Is were broad with a FWHM between 100 and 300 km/s. They were slightly blue-shifted and nearly symmetric. The average maximum velocities observed were -282 km/s and -240 km/s respectively for the red and blue wings. When they compared these results to predictions made by the models they concluded that the infall models produced lines that were too narrow and  wings that extended to velocities that were too small. IPC profiles on the other hand must arise largely from infalling material, for two reasons. Firstly the red-shifted absorption feature must be formed in infalling material as it is located at velocities of the order of the free-fall velocity at a few stellar radii. The second reason stems from a trend described by the authors, associating the amount of emission seen in IPC profiles with accretion rates.

So it is clear then that as with the Balmer series, the  major discrepancy between computed profiles and observations in the case of the Pa$\beta$ line, lies in failure of studies based on the MAM to predict Type I profiles. In an attempt to account for this fact the authors considered several ideas. Firstly the systems geometry may play a role in determining the shape of the line profiles. The MAM  predicts red-shifted absorption for systems observed at high inclinations. The authors looked at the distribution of inclinations for a sample of the stars whose inclinations were known. They noted no significant  difference between the distribution of Type I and IPC sources. However this result was obtained from a small sample of stars and should be considered with extreme caution. Secondly they suggested that it may be possible that the presence of a red-shifted absorption feature is related to the temperature of the accretion shock. This temperature is not however known accurately enough in order to be able to investigate this idea. Lastly according to \cite{Muzerolle98(a)} if HI emission originates in infalling gas then the magnetospheric accretion model predicts a correlation between emission line strength and accretion rate. \cite{Folha01} find no such agreement for Type I profiles but there does seem to be a trend for IPC profiles. Stars with larger accretion rates tend to have a larger EW in the emission component of the IPC Pa$\beta$ line profile.

Looking at our data in the same manner as \cite{Folha01} we see that both sets of results agree well. DG Tau, V536 Aql and LkH$\alpha$ 321 all have Type I Pa$\beta$ profiles.  DG Tau is actually slightly red-shifted with a radial velocity of$\sim$ +8 km/s. It has a FWMH of $\sim$225 km/s and a prominent blue-wing that extends to approximately -380 km/s.  V536 Aql is blue-shifted to a velocity of $\sim$-30 km/s, it has a FWHM of $\sim$167 km/s and a red wing that extends to $\sim$350 km/s and also a blue-wing to $\sim$ -250 km/s. The LkH$\alpha$ 321 Pa$\beta$ line profile is also a Type I profile. It is perhaps the most symmetric of the three. It is blue-shifted to $\sim$ -2 km/s and its red and blue wings extend to approximately 300 km/s. It FWHM is $\sim$220 km/s. RW Aur has a Type IV R profile. This line emission was also classified as an IPC profile by Folha and Emerson even though the red shifted absorption feature is very weak. The RW Aur Pa$\beta$ line has a FWHM of -320 km/s and its blue wing extends to -400 km/s.

\subsection{Spectro-astrometry}
As stated an important part of our work was to produce specto-astrometric plots of the emission from our sources. Spectro-astrometry was used by \cite{Hirth97} where they presented the optical FEL profiles of 38 T-Tauri stars. This technique also played an important role in the recent interesting studies by \cite{Takami01} and \cite{Takami03}. For example \cite{Takami01} presented a study of the circumstellar structure of RU Lupi. They noted that the H$\alpha$ profile from this source had extended line wings and that emission in these wings was offset from the star by 20-30 mas. They used spectro-astrometric plots to present this data and compared them with plots made of the [SII]$\lambda\lambda$ 6716, 6731 and [OI]$\lambda$ 6300 lines. The same technique was used by them in \cite{Takami03} to study CS Cha. The technique of spectro-astrometry was also used to great effect in our [FeII] paper (DWRC) where we were able to compare the peak velocity and maximum offset of [FeII] emission from many sources with molecular hydrogen observations of the same objects \citep{Davis01}. We will repeat this approach here but this time we compare our Pa$\beta$ results with our optical forbidden emission line results. In the case of DG Tau, LkH$\alpha$321 and  RW Aur [FeII] emission is also considered.
  
We shall now look at each source separately. Table 2 compares the peak velocities and maximum offsets measured for each source in all the different lines. In the case of the FELs peak velocities can be estimated from gaussian fits to the emission line profiles and from the spectro-astrometric plots. For Pa$\beta$ we can only rely on our spectro-astrometric plots to estimate the peak velocities at which we are seeing the outflowing material, due to the fact that we only measure a positional displacement in the line wings.  
\subsubsection{DG Tau}
Figure 5(a)(ii) shows the plot of the offsets against velocity that were measured for the [FeII] line (DWRC). We mainly detected [FeII] emission from the high velocity jet and the plot shows that emission reaches a maximum of approximately -280 km/s and extends to approximately 6$\arcsec$. The [FeII] emission line profile shows that emission peaks at $\sim$-235 km/s. In the optical FELs we detect [SII]$\lambda$6731 and [OI]$\lambda$6300 emission from the source. We also made spectro-astrometric plots of these lines. From the line profile we see that [SII]$\lambda$6731 emission has a low velocity peak at -47 km/s and a HVC at -240 km/s. From the plot the low velocity emission reaches an offset of $\sim$ 0$\farcs$6 and we see a peak in velocity in the blue  at $\sim$ -220 km/s and $\sim$ 1$\farcs$4. The [OI]$\lambda$6300 emission from this source is double-peaked. We see a LVC at -47 km/s and 0$\farcs25$ and a HVC at -266 km/s and 1$\arcsec$. 

From Figure 5(a)(i) it is clear that the Pa$\beta$ emission is extended. We measure offsets in the blue wing out to almost 0$\farcs$5. Emission is seen to peak at $\sim$-250 km/s and extend to $\sim$-300 km/s at most. Through a comparison between the spectro-astrometric plots for the FELs and that for Pa$\beta$ we can conclude that Pa$\beta$ is tracing the high velocity emission from the DG Tau outflow. The displacement of the blue-wing aligns well with positional displacements measured in the FELs i.e. offsets are in the same direction and peak at velocities that compare well in each case. The fact that the outflow is traced by Pa$\beta$ and the FELs at very similar velocities suggests that the lines are excited in the same pocket of gas. We also see that for all lines except [FeII] there is an initial increase in velocity with increasing distance a feature which is typical of magnetically driven flows \citep{Lago84, DeCampli81}. Increasing velocity with increasing distance is not seen in the case of [FeII] as the [FeII] emission peaks much further out where the jet is presumably interacting more with the ambient medium and for example forming bow shocks. The shock geometry may effect the velocity dependent offsets.   
\subsubsection{V536 Aql}
We detected no [FeII] emission from V536 Aql but we did detect optical forbidden emission. Figure 5(b) shows the offset plot for the [SII]$\lambda$6731 and [OI]$\lambda$6300 lines. The [SII]$\lambda$6731 line profile is double peaked with a HVC at approximately $\sim$-84 km/s and a LVC at $\sim$-20 km/s. This is clearly seen in the [SII]$\lambda$6731 emission line profile for this source, Figure 2(c). Offsets from this line reach $\sim$0$\farcs9$ for the HVC and approximately 0$\farcs$15 for the low velocity emission. The [OI]$\lambda$6300 line profile shows a strong LVC at -10 km/s and some evidence for high velocity emission at $\sim$-65 km/s. Figure 5(b)(iv) shows this high velocity emission to peak at a displacement of $\sim$0.85 $\arcsec$ and offsets reach $\sim$0$\farcs$05 at the value of the low velocity emission. 

Looking at the Pa$\beta$ we see the line profile is symmetric but the spatial offsets in the wings are in the opposite sense and in a direction consistent with outflowing gas, traced by the optical FELs. From the spectro-astrometric plot we see high velocity emission  at $\sim$-120 km/s displaced to a distance of $\sim$ 0$\farcs$08. We see low velocity emission at $\sim$+40 km/s displaced to a maximum distance of $\sim$ 0$\farcs$1. Hence offsets again peak {\bf within} the same velocity range and with the same orientation. We are seeing a HVC in Pa$\beta$ at -120 km/s which we also see in the [SII]$\lambda$6731 and [OI]$\lambda$ 6300 lines. It is also clear that velocity increases initially with distance from the star and so we again conclude that the Pa$\beta$ emission from V536 Aql is tracing the bipolar outflow from this source and that emission from this outflow makes up the wings of the emission line profile. 

As stated above we also have spectro-astrometric results for the H$\alpha$ emission from this source. We detect very strong H$\alpha$ close to source and very weak emission from the jet. Again we see a bipolar displacement that is symmetric. The blue-shifted emission reaches a maximum velocity of approximately -270 km/s at a distance of 0$\farcs13$. The red-shifted emission reaches a velocity of $\sim$260 km/s at 0$\farcs$07.   
  
\subsubsection{LkH$\alpha$ 321}
We see weak [FeII] emission towards this star. It was also very weak in the optical FELs. We only detected the [SII]$\lambda$6731 and [OI]$\lambda$6300 lines close to the source. For [SII]$\lambda$6731 (see Figure 6(a)(ii))the emission peaks at about -340 km/s and is measured out to $\sim$0$\farcs$25 at most. In the case of the [OI]$\lambda$6300 (see Figure6(a)(iii)) emission peaks at -412 km/s and 0$\farcs$17. It should be noted that such small offsets are not typical of the HVCs of T-Tauri outflows as seen in optical FELs. Also velocities are not typically as high as we see them here. These results suggest that the outflow from LkH$\alpha$321 may be pointing directly towards us.  

Figure 6(a)(i) shows the spectro-astrometric plot of the Pa$\beta$ emission. This profile has extended blue and red wings and as we can see the Pa$\beta$ traces both poles of the outflow. Velocity peaks at around -160 km/s in the blue lobe and extends to 1$\farcs$3 at the most. Velocity in the red-wing peaks at approximately 120 km/s and a distance of less than 0$\farcs$1. Although the outflow from this YSO has been imaged \cite{Mundt98} there is no spectroscopic data available in the literature which clarifies the orientation of the red and blue lobes. Therefore our [SII]$\lambda$6731 and [OI]$\lambda$6300 spectro-astrometric plots are the first results which give this orientation. The offsets we measure in Pa$\beta$ are in the same direction as those measured in the FELs. Again there is an initial velocity increases with offset. In this case we do not see a strong agreement between peak velocities, however it is not necessary for the Pa$\beta$ emission to be formed in the same region of the outflow as the optical FELs.
\subsubsection{RW Aur}
RW Aur is the only source for which we do not have new optical data. However it was covered by Hirth et al 1997 paper so we can use their findings as a reference. Results published in our [FeII] paper showed that [FeII] emission in both lobes at peak  LSR velocities of $\sim$ -175 km/s and $\sim$ 150 km/s. The corresponding spectro-astrometric plot (previously published in DWRC; also see Figure 6(b)(ii)) shows maximum displacements of $\sim$ 1$\arcsec$ in both lobes. Hirth et al 1997, presented spectro-astrometric plots of the[SII]$\lambda$6731 line and published values of peak velocity of -166 km/s and 99 km/s for the red and blue lobes respectively. 

In Pa$\beta$ we measured no offsets and this is the only source in our sample showing this behaviour. The spectro-astrometric plot Figure 6(b)(i) shows the Pa$\beta$ emission to be coincident with the source. This is an interesting result as the Pa$\beta$ line profile is an inverse P Cygni profile and as we discussed above in connection with the results of \cite{Folha01} and others, this type of profile is well explained by the MAM. The IPC profile for this source shown in Figure 1(d) has a blue-wing that reaches a maximum velocity of $\sim$ -400 km/s however we see no positional displacement in this emission region.

\begin{table*} 
\begin{tabular}{llllll}
\hline \hline           
Object          &Emission Line               &$V_{Blue}(km/s)$      &Offset(arcseconds)       &$V_{Red}(km/s)$ &Offset(arcseconds)  
\\ \hline
DG Tau          &[FeII] 1.644$\mu$m          &-235                  &$\sim$6                  &                &
\\ 
                &[SII]$\lambda$6731                  &-47 (L); -240 (H)     &$\sim$0.6; $\sim$1.4     &                &
\\ 
                &[OI]$\lambda$6300                   &-47 (L); -266 (H)     &$\sim$0.25; $\sim$1      &                &
\\
                &Pa$\beta$                   & -250                 &$\sim$0.5                &                &
\\ 
V536 Aql        &[SII]$\lambda$6731                  &-20 (L); -84 (H)     &$\sim$0.15; $\sim$0.9                &                &
\\ 
                &[OI]$\lambda$6300                   &-10 (L); -65 (H)     & $\sim$0.05; $\sim$0.85             &                &
\\ 
                &Pa$\beta$    &-120          &$\sim$0.08               &40              &$\sim$-0.1
\\ 
                &H$\alpha$    &-270          &$\sim$0.13               &260              &$\sim$-0.07
\\ 
LkH$\alpha$ 321 &[FeII] 1.644$\mu$m          &                      &                         &229             &
\\                                                                                                  
                &[SII]$\lambda$6731                  &-340                  &$\sim$0.25               &                &
\\
                &[OI]$\lambda$6300                   &-412                  &$\sim$0.17               &                &
\\       
                &Pa$\beta$                   &-160                  &$\sim$1.3               &120             &$\sim$-0.075
\\ 
RW Aur          &[FeII] 1.644$\mu$m          &-175                  &$\sim$1                  &150             &$\sim$-1
\\                 
                &Pa$\beta$                   &0                     &0                        &0               &0
\\                                                                                         
                &[SII]$\lambda$6731*         &-166                  &                         &99              &
\\ \hline  
\end{tabular}
\caption{Table comparing peak LSR velocities and offsets measured for each source in Pa$\beta$ , [SII]$\lambda$6731, [OI]$\lambda$6300 or [FeII]. * The [SII]$\lambda$6731 velocity measurements for RW Aur were taken from Hirth et al 1997.}
\end{table*}

\subsection{Common Trends}
\subsubsection{Circumstellar disk gaps}

Our results show that for three out of four of the TTSs studied by us, the positional displacement measured in the Pa$\beta$ line wings is consistent with outflowing material. Central to our argument is the comparison with the optical forbidden emission from these sources. One notable feature of the FELs of T-Tauri stars is that they are generally blue-shifted. It is known that the red-shifted part of the flow is often obscured by the optically thick circumstellar disks that have been proven to exist around these stars \citep{ed93}. Hence the fact that for two out of four of our sources namely V536 Aql and LkH$\alpha$ 321 we detect Pa$\beta$ emission from both sides of the outflow, is very interesting.

As mentioned above similar spectro-astrometric results were published for the H$\alpha$ emission from RU Lupi and CS Cha by \cite{Takami01} and \cite{Takami03}. They also measured a bipolar displacement in the H$\alpha$ emission which they suggested provided further evidence in favour of the existence of gaps in the circumstellar disks of young stars. If a gap existed it would allow the red-shifted emission to be seen. We should think of these holes as areas where dust has been cleared (and not necessarily gas), hence the opacity of the circumstellar material in this region of the disk has been lowered. It is a likely scenario that these gaps occur in more evolved T-Tauri stars where the process of planet formation has begun and the dust grains have started to grow \citep{Testi03, Shuping03, Wood02}.  

To date the existence of holes in circumstellar disks has mainly been inferred from a study of the spectral energy distributions (SEDs) of young stars. The fact that the SEDs of many young stars show a dip at mid-infrared wavelengths or often exhibit a small or non-existence NIR excess has been taken to prove the presence of regions of cleared circumstellar material \citep{jensen97, marsh92,mathieu91}. It is very tempting to then make the inference that these regions of cleared material suggest the presence of a protoplanetary or planetary object \citep{Stein03, Calvet02, Bryden00}.

In the case of RU Lupi \cite{Takami01} measured a displacement of 20-30 mas which they claim corresponds to a gap radius of 3-4AU. To further prove their point \cite{Takami01} plotted the SED of RU Lupi. They measured a shallow dip at 4-15$\mu$m which corresponds to temperatures of 200-900K and the radiative temperature at a radius of 0.1-2AU from the star. This fits well with the estimate of a gap radius of 3-4AU taken from the extent of the displacement measured in the H$\alpha$ wings.     

We will now consider the possibility that the presence of a gap in the circumstellar disk around V536 Aql is allowing us to detect the red-shifted Pa$\beta$ emission. As stated V536 Aql is a binary star and circumstellar material has been detected around this source. We measure a displacement in the red-wing of $\sim$0$\farcs$1. V536 Aql is at a distance of 200pc and this suggests a gap radius of 20 AU.  

A SED for this source was available to us through a private communication (see Gras-Vel\'{a}zquez et al 2003 in preparation). Similar to what is observed for RU Lupi we see a dip in emission between 2$\mu$m and 16$\mu$m indicating a lack of material in the temperature range 180-1450 K. A value of 4.5L$\sun$ is estimated by Gras Vel\'{a}zquez et al for this source. The radial temperature dependence of the disk is given by T(r)=$(2/3\pi)^{1/4}$$T_{*}$$(r/R_{*})^{-q}$, the constant q depends on the spectral index $\beta$ for the dust opacity law \citep{Beck90}. We choose a value of $\beta$ equal to 1 as it is believed that a value of $\beta$ $\leq$ 1 describes the particle emissivity in circumstellar disks \citep{Beck91}.  We estimate a gap radius of $\sim$18 AU. This value agrees well with what we calculated above using our spectro-astrometric data. 

As well as seeing red-shifted Pa$\beta$ emission we also see red-shifted H$\alpha$ emission from V536 Aql. Hence we can also estimate the gap size from our spectro-astrometric results for the H$\alpha$ line. The red-shifted emission is displaced out to distance of 0\farcs07 which corresponds to a gap radius of $\sim$14AU. The blue-shifted emission is offset further than the red, out to a distance of 0$\farcs$3. Overall both the SED and the spectro-astrometric data we have for V536 Aql points to the existence of a region in the disk where the opacity of the circumstellar material is low enough for red-shifted emission from both Pa$\beta$ and H$\alpha$ to be detected. The fact that we measure a bigger gap in Pa$\beta$ than in H$\alpha$ is expected as the transparency of the disk will be greater at longer wavelengths. Hence it should be possible to use these lines in conjunction with other HI lines of known intensity as a diagnostic of the transparency of the circumstellar disk. 

\begin{figure}
\epsfig{file=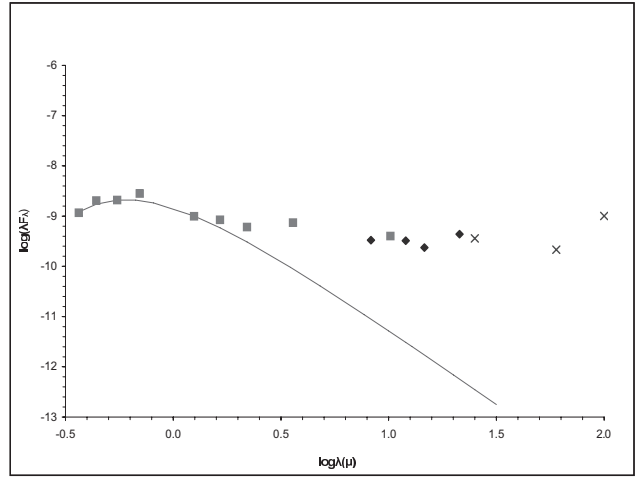,width=8.5cm,height=6.5cm}
\caption{SED of LkH$\alpha$ 321. A blackbody curve representing the stellar photosphere is fitted to the flux at V (0.55$\mu$m). The square boxes represent data points for the U to N bands, the diamonds represent measurements taken in the bands used by the MSX and the crosses are the IRAS points}
\end{figure} 

As stated we see bipolar Pa$\beta$ emission for two out of four of our sources. The second of these sources is LkH$\alpha$ 321. We measure offsets in the Pa$\beta$ red-wing out to $\sim$0$\farcs075$. This source is at a distance of 550pc hence this positional displacement corresponds to a gap size of 40AU. We plotted the SED of this source from 0.354$\mu$m (U) to 100$\mu$m and the photometric measurements in this range were obtained from a variety of sources. The flux measurements in the U, B, V and R bands were taken from \cite{Herbst99} and in the L and N bands from \cite{Cohen74}. The 2Mass sky survey provided the photometry in the J, H and K bands. We also accessed the Midcourse Space Experiment (MSX) archive to obtain flux measurements at 4.2, 8.28, 12.13, 14.65 and 21.34 microns. Lastly measurements at 12$\mu$m, 25$\mu$m, 60$\mu$m and 100$\mu$m were extracted from the Infrared Astronomical Satellite (IRAS) point source catalogue.

The clear infrared excess in the SED of LkH$\alpha$ 321 points to the existence of a circumstellar disk however we see no evidence for a gap in this disk. This is surprising considering the good agreement between the specto-astometry and the dip in the SED we of V536 Aql. \cite{Chav81} give the luminosity of this source at $\sim$146$\sun$. Considering this value we would expect a gap of 40 AU to correspond to a temperature of $\sim$ 440K or a wavelength of $\sim$ 6$\mu$m in the SED. We were unable to obtain photometric data for this source in the M band i.e. between about 4 and 8 microns, therefore it may be possible that there is a lack of emission in this range which we have not yet been able to confirm. 

For DG Tau we measure a displacement in the blue-wing only. DG Tau appears less evolved than either V536 Aql or LkH$\alpha$ 321. This is clearly seen by the fact that it drives more powerful outflows and hence must have a greater rate of accretion. See Figures 3 and 4 for a comparison between the jets from DG Tau and V536 Aql, traced in the optical FEL's of [OI]$\lambda$6300 and [SII]$\lambda$6731. Also DG Tau is known to have a flat SED \citep{Adams87, Lodato01}. Hence the SED of DG Tau shows no evidence of a disk gap or hole which would allow the red-shifted jet to be traced in Pa$\beta$. This ties in well with our results for V536 Aql. If the lower opacity which allows us to see the red-shifted Pa$\beta$ emission is as a result of a growth in the dust grains in the inner regions of the circumstellar disk or clearing of dust due to the presence of a planet, then it would be reasonable to expect the circumstellar disk of a less evolved YSO like DG Tau to have no such regions of low opacity. 

\subsubsection{Acceleration in different lines}  

Previous spectro-astrometric studies \citep{Hirth97} have shown that amongst the individual optical FELs displacement increases with decreasing density. For example the displacement at high velocities has been shown to be larger in [SII] than in [OI] and this suggests decreasing densities as we move away from the star \citep{Hamann94}. We can now extend this argument from the optical into the near-infrared. 

Our results show that offset increases as we move from Pa$\beta$ to the optical FELs and then to [FeII]. For example for DG Tau we trace the outflow in [FeII] out to 6$\arcsec$, in [SII]$\lambda$6731 out to 1$\farcs$4, in [OI]$\lambda$6300 out to 1$\arcsec$ and finally in Pa$\beta$ out to 0$\farcs$5. In V536 Aql displacement decreases for the emission lines in the same order as above, from 0$\farcs$9 to 0$\farcs$85 to 0$\farcs$08 (we detected no [FeII] emission from V536 Aql). Lastly for LkH$\alpha$321 offsets decrease from 0$\farcs$25 to 0$\farcs$17 to 0$\farcs$075. We only detect very weak [FeII] emission from the LkH$\alpha$321 jet so it was not possible to measure a positional displacement.

\cite{Hamann94} gives the temperatures and densities of the [OI] and [SII] emission regions as 9000$\leq$ $T_{e}$ $\geq$14000K,  $n_{e}$ $\leq$ 5 $\times$ $10^{5}$ to $\sim$ $10^{7}$ $cm^{-3}$ and $T_{e}$ $\geq$13,000K, $10^{3}$ $\leq$ $n_{e}$ $\leq7$ $\times$ $10^{4}$ $cm^{-3}$ respectively. For atomic hydrogen lines like Pa$\beta$ it is believed that they are formed under similar temperatures as the optical FELs but much higher densities. \cite{KwanA88} estimate the density of the hydrogen line emission zone at $10^{10}$ to $\sim$ $10^{12}$ $cm^{-3}$. \cite{Muzb} modelled Na D and hydrogen lines to constrain the physical parameters of the magnetospheric infall zone and they estimated gas temperatures in this region of about 10,000K (in the case of DP Tau). It is also clear from our spectro-astrometric plots that Pa$\beta$ is formed very close to the central star. Lastly [FeII] emission derives from gas at a density of $n_{e}$ $\sim$ $10^{4}$ $cm^{-3}$ and a temperature of the order of $10^{4}$K \citep{Hollenbach89}. 

This comparison of the displacement in the different emission lines clearly reflects the fact that density decreases as we move away from the driving source of the outflows. Pa$\beta$ forms in the regions very close to the star where temperatures and densities are high and where one is observing both accretion and outflow. In fact our results suggest that the bulk of the Pa$\beta$ emission forms in the hot accretion zones as we find that only the emission making up the wings shows spatial offsets. The optical FELs form further out where temperatures are similar but densitites are lower. Lastly the [FeII] traces the outflow out to distances where densities are much lower again.

\section{Conclusion}
 To summarise, we present the results of a study where we looked at Pa$\beta$ emission from four T-Tauri stars and through the technique of spectro-astrometry we examined the resultant data for evidence of the presence of outflows. We then directly compared our results with optical forbidden emission line spectra in which we can clearly detect the presence of outflows. From this study we come to the following conclusions:

1. The HI Pa$\beta$ line profiles of three out of four of the sources are all Type I profiles in accordance with the classification scheme of \cite{Reipurth96}. The MAM which is the basis for theories describing how atomic hydrogen emission lines form in accretion zones connected with YSOs, cannot account for Type I profiles. In particular it cannot explain the presence of the line wings we see in these sources.

2. We measure positional displacements, in the Paschen beta emission which makes up the wings of the line profiles in all of the three sources with Type I profiles.

3. In all of these sources the velocities at which we measure the offsets are comparable to the velocities at which we see the outflows in the [SII]$\lambda$6731 and [OI]$\lambda$6300 emission.

4. In all of these sources offset is seen to initially increase with velocity a trend which is typical of magnetically driven outflows. The sense of the outflows (i.e. blue or red-shifted) reflected in the  Pa$\beta$ plots matches spatially that which we see in the optical FELs. Plus where we see both lobes of the flow e.g in the case of V536 Aql they are symmetric.

5. Our fourth source namely RW Aur has an Inverse P Cygni profile which is best explained by the MAM. We measure no positional displacements in the Pa$\beta$ emission from RW Aur.

6. We suggest that the bipolar nature of the Pa$\beta$ and H$\alpha$ emission from V536 Aql may be indicative of the presence of a gap or region of low opacity in the circumstellar disk of this source. This is also suggested by the SED of V536 Aql. We also measure a bipolar displacment in Pa$\beta$ emission from LkH$\alpha$ 321 however the SED we plotted for this source did not confirm the presence of a gap. As stated if Pa$\beta$ can be used to study outflows it will do so on very small spatial scales. Our results confirm this and the fact that we may be able to use this line to detect the presence of gaps in circumstellar disks demonstrates how useful this line will be. Interestingly the comparison between the H$\alpha$ and Pa$\beta$ lines for V536 Aql suggests that we should be able to use several HI lines of known intensity to construct a profile of the transparency of the disk. Lastly we also observe how the extent to which the different emission lines trace the outflow, increases as we move to lower temperatures and densities.      

It is fair to say that we have presented a strong argument in favour of line wings being formed in areas of outflow. An extensive study of TTSs and Class I sources is planned to add further weight to this argument. As the Class I sources are known to drive more powerful outflows than the T-Tauri stars it will be interesting to see if this is reflected in the Pa$\beta$ emission. Other studies such as that by \cite{Takami01} mentioned above back up our conclusions.
 
\acknowledgements {We would like to thank Watson Varicatt for his assistance at the telescope during the 2001 observing run at UKIRT and the staff of UKIRT and the INT who were responsible for the service observations. The UKIRT is operated by the Joint Astronomy Centre on behalf of the U.K. Particle Physics and Astronomy Research Council. The INT is operated on the island of La Palma by the Isaac Newton Group in the Spanish Observatorio del Roque de los Muchachos of the Instituto de Astrofisica de Canarias. This research has made use of the SIMBAD database, operated at CDS, Strasbourg, France. We would also like to thank Agueda Gras-Vel\'{a}zquez for providing us with the information on the SED of V536 Aql.}

\end{document}